\documentclass[aps,amsmath,amssymb,superscriptaddress]{revtex4}
\usepackage{graphicx}

\usepackage{color,soul}

\newcommand{\be}{\begin{equation}}
\newcommand{\ee}{\end{equation}}
\newcommand{\bea}{\begin{eqnarray}}
\newcommand{\eea}{\end{eqnarray}}
\newcommand{\ba}{\begin{eqnarray*}}
\newcommand{\ea}{\end{eqnarray*}}

\newcommand{\dis}{\displaystyle}

\newcommand{\up}{\uparrow}
\newcommand{\down}{\downarrow}
\newcommand{\fract}[2]{\frac{\dis #1}{\dis #2}}

\newcommand{\bw}{\begin{widetext}}
\newcommand{\ew}{\end{widetext}}

\begin{document}
\title{Mottness at finite doping and charge-instabilities in cuprates}

\author{S. Peli}
\affiliation{Department of Physics, Universit\`a Cattolica del Sacro Cuore, Brescia I-25121, Italy}
\affiliation{Department of Physics, Universit\`a degli Studi di Milano, Italy}

\author{S. Dal Conte}
\affiliation{IFN-CNR, Dipartimento di Fisica, Politecnico di Milano, 20133 Milano, Italy}

\author{R. Comin}
\affiliation{Quantum Matter Institute, University of British Columbia, Vancouver, BC V6T 1Z4, Canada}
\affiliation{Department of Physics and Astronomy, University of British Columbia, Vancouver, BC V6T 1Z1, Canada}
\affiliation{Present address: Dept. of Physics, Massachusetts Institute of Technology, Cambridge, MA 02139-4307 (US)}

\author{N. Nembrini}
\affiliation{Department of Physics, Universit\`a Cattolica del Sacro Cuore, Brescia I-25121, Italy}
\affiliation{Department of Physics, Universit\`a degli Studi di Milano, Italy}

\author{A. Ronchi}
\affiliation{Department of Physics, Universit\`a Cattolica del Sacro Cuore, Brescia I-25121, Italy}
\affiliation{Department of Physics and Astronomy, KU Leuven, Celestijnenlaan 200D, B-3001 Heverlee, Leuven, Belgium}

\author{P. Abrami}
\affiliation{Department of Physics, Universit\`a Cattolica del Sacro Cuore, Brescia I-25121, Italy}

\author{F. Banfi}
\affiliation{Department of Physics, Universit\`a Cattolica del Sacro Cuore, Brescia I-25121, Italy}
\affiliation{i-LAMP (Interdisciplinary Laboratories for Advanced Materials Physics), Universit\`a Cattolica del Sacro Cuore, Brescia I-25121, Italy}

\author{G. Ferrini}
\affiliation{Department of Physics, Universit\`a Cattolica del Sacro Cuore, Brescia I-25121, Italy}
\affiliation{i-LAMP (Interdisciplinary Laboratories for Advanced Materials Physics), Universit\`a Cattolica del Sacro Cuore, Brescia I-25121, Italy}

\author{D. Brida}
\affiliation{Department of Physics and Center for Applied Photonics, University of Konstanz, 78457 Konstanz, Germany}

\author{S. Lupi}
\affiliation{CNR-IOM Dipartimento di Fisica, Universit\`a di Roma La Sapienza P.le Aldo Moro 2, 00185 Rome, Italy}

\author{M. Fabrizio}
\affiliation{Scuola Internazionale Superiore di Studi Avanzati (SISSA) and CNR-IOM Democritos National Simulation Center, Via Bonomea 265, 34136 Trieste (Italy)}

\author{A. Damascelli}
\affiliation{Quantum Matter Institute, University of British Columbia, Vancouver, BC V6T 1Z4, Canada}
\affiliation{Department of Physics and Astronomy, University of British Columbia, Vancouver, BC V6T 1Z1, Canada}

\author{M. Capone}
\affiliation{Scuola Internazionale Superiore di Studi Avanzati (SISSA) and CNR-IOM Democritos National Simulation Center, Via Bonomea 265, 34136 Trieste (Italy)}

\author{G. Cerullo}
\affiliation{IFN-CNR, Dipartimento di Fisica, Politecnico di Milano, 20133 Milano, Italy}

\author{C. Giannetti}
\affiliation{Department of Physics, Universit\`a Cattolica del Sacro Cuore, Brescia I-25121, Italy}
\affiliation{i-LAMP (Interdisciplinary Laboratories for Advanced Materials Physics), Universit\`a Cattolica del Sacro Cuore, Brescia I-25121, Italy}

\begin{abstract}

The intrinsic instability of underdoped copper oxides towards inhomogeneous states is one of the central puzzles of the physics of correlated materials. %This fragility manifests itself in a variety of ways including the spontaneous emergence of charge-order patterns at low temperature. %The charge ordering phenomenon appears below a characteristic temperature and only up to a critical hole doping ($p_{cr}$) that is very close to the optimal doping concentration for superconductivity and constitutes the turning point for many other properties that suggest the presence of a zero temperature quantum critical point (QCP).
%The experimental evidences show a general trend where various instabilities, that break different symmetries at the nanometric scale, appear below a characteristic temperature and only up to a critical hole doping $p_{cr}$. 
%The universality of these findings revive the long debated question whether the spontaneous development of charge inhomogeneities or other possible orders responsible for the QCP are the just result of specific instabilities at the Fermi level or they are the low-energy manifestation of a more general precursory state, which arises from strong electronic correlations suddenly changing at $p_{cr}$.
The influence of the Mott physics on the doping-temperature phase diagram of copper oxides represents a major issue that is subject of intense theoretical and experimental effort.
Here, we investigate the ultrafast electron dynamics in prototypical single-layer Bi-based cuprates at the energy scale of the O-2\textit{p}$\rightarrow$Cu-3\textit{d} charge-transfer (CT) process. We demonstrate a clear evolution of the CT excitations from incoherent and localized, as in a Mott insulator, to coherent and delocalized, as in a conventional metal. This reorganization of the high-energy degrees of freedom  occurs at the critical doping $p_{cr}\approx$0.16 irrespective of the temperature, and it can be well described by dynamical mean field theory calculations. We argue that the onset of the low-temperature charge instabilities is the low-energy manifestation of the underlying Mottness that characterizes the $p<p_{cr}$ region of the phase diagram. This discovery sets a new framework for theories of charge order and low-temperature phases in underdoped copper oxides.
\end{abstract}
\maketitle

When charge carriers are chemically doped into a Mott or charge-transfer insulator, the system progressively evolves into a metal whose electronic properties are strongly reminiscent of the on-site electronic correlations\cite{Lee2006}. In the case of copper oxides, the complexity of the problem has roots in the intertwining between the high energy scale of the Mott physics\cite{Phillips2006} (several electronvolts) and the low-energy phenomena that typically emerge in the low temperature/doping region of the phase diagram\cite{Abbamonte2005,Fradkin2012,Alloul2014,Keimer2015}. For example, the vicinity to the Mott insulating phase at zero doping ($p$=0) has been advocated\cite{Emery1993,Castellani1995,Kivelson1998} as the main mechanism that drives the freezing of the charge carriers within the CuO$_2$ unit cell and the reduction of their kinetic energy, thus facilitating the low-temperature formation of charge-ordered states and other forms of order that spontaneously break the translational symmetry of the underlying crystal. In fact, the universal tendency to develop short-ranged incommensurate charge density waves (CDW) in the underdoped region of the phase diagram and below a characteristic temperature has been recently reported in both hole- and electron-doped copper oxides by X-ray diffraction\cite{Ghiringhelli2012,Achkar2012,Chang2012,BlancoCanosa2014,Tabis2014,daSilvaNeto2015,Comin2015}, tunneling microscopy\cite{Comin2014,daSilvaNeto2014} and nuclear magnetic resonance\cite{Wu2015}. More in general, the breaking of the rotational symmetry from $C_4$ to $C_2$ (nematicity) has been argued from X-ray and neutron scattering experiments\cite{Zimmermann1998,Hinkov2007,Tranquada2008} and directly imaged by scanning tunneling microscopy (STM)\cite{Lawler2010}. The signature of intra-unit-cell magnetic order has been observed by neutron scattering\cite{Li2010} and Kerr effect measurements\cite{Karapetyan2012}.

In the case of multiband systems, such as cuprates, the oxygen bands play a fundamental role in renormalizing the energy scale at which the Mott physics can be studied. Considering the simplest case of the parent insulator ($p$=0), the valence fluctuations of Cu-3$d^9$ are suppressed by the strong Coulomb repulsion ($U_{dd}\sim10$ eV) between two electrons occupying the same Cu orbital. The lowest excitation is thus the charge-transfer (CT) of a localized Cu-3$d_{x^2-y^2}$ hole to its neighbouring O-2$p_{x,y}$ orbitals (see Fig. 1a,b), with an energy cost $\Delta_{CT}\sim$2 eV$<U_{dd}$. In the optical conductivity, this process is revealed by a typical CT edge at $\hbar\omega$=$\Delta_{CT}$, which corresponds to the onset of optical absorption by particle-hole excitations in the complete absence of a Drude response\cite{Uchida1991}. Since conventional spectroscopic techniques probe the physical properties at equilibrium, in which only the fluctuations at the energy scale $k_BT \ll \Delta_{CT}$ are thermally activated, the relation between the low-temperature onset of symmetry-breaking instabilities and the Mott physics involving energy scales of the order of $\Delta_{CT}$ has remained hitherto unexplored.

\begin{figure}
\includegraphics[keepaspectratio,clip,width=1\textwidth]{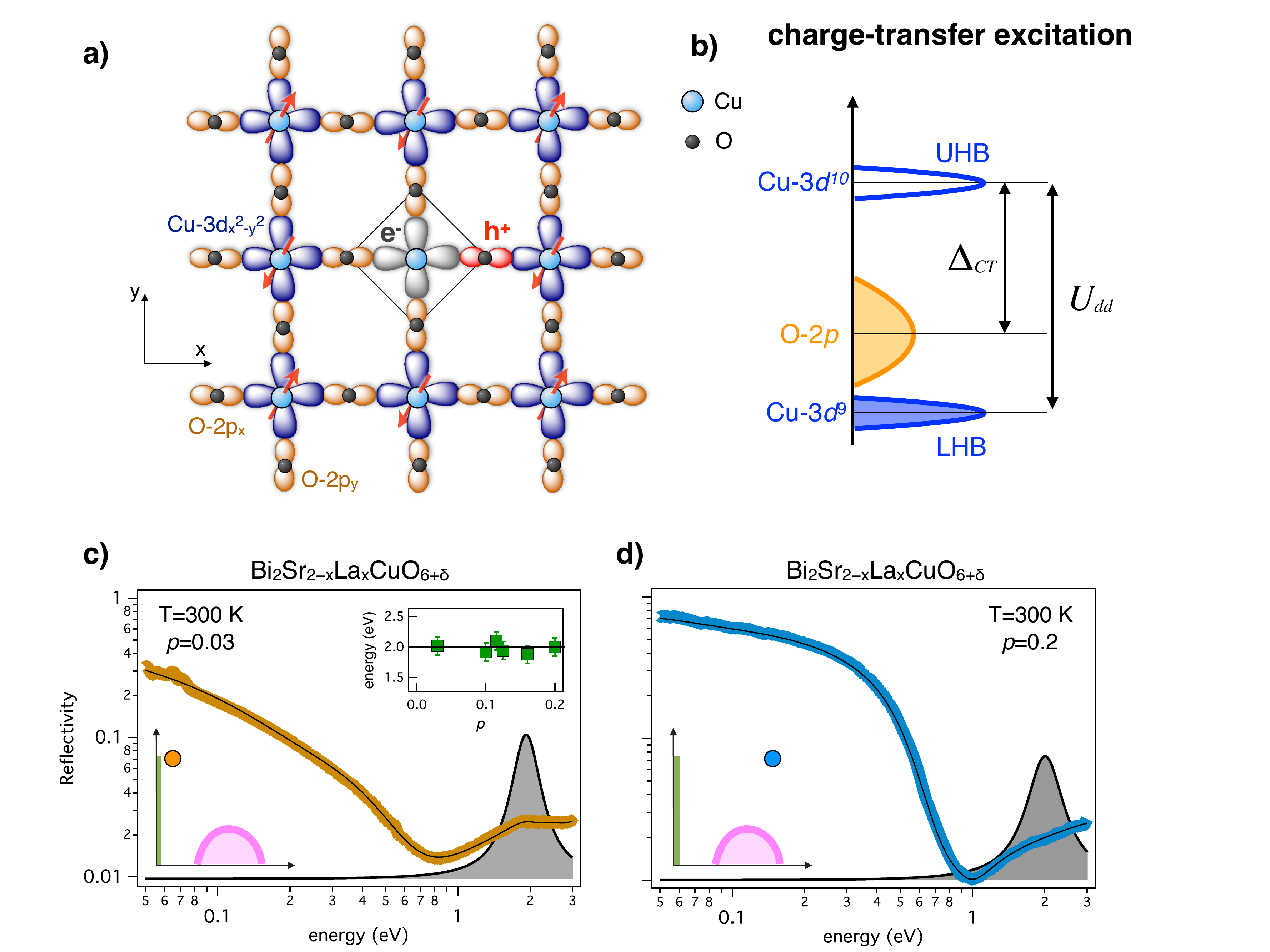}
\caption{\textbf{Charge-transfer excitation and optical properties of cuprates.} \textbf{a,b)} A sketch of the generic charge-transfer process in the Cu-O layer of copper oxides is shown. The upper (UHB) and lower (LHB) Hubbard bands, corresponding to the Cu-3$d^{10}$ and Cu-3$d^{9}$ configurations, are indicated \textbf{c)} The reflectivity curve, $R(\omega)$ of the underdoped sample is reported (yellow dots). The black line is the fit to the data obtained from a model dielectric function which contains an extended-Drude term and three Lorentz oscillators that account for the high-energy transitions. The contribution of the first interband oscillator ($\omega_{CT}$), attributed to the charge-transfer process, to the total dielectric function is reported as a grey region. The inset displays $\omega_{CT}$ as a function of the hole concentration. The left-bottom inset shows the position of the sample in the $p$-$T$ phase diagram. The pink line represents the superconducting dome, while the green line indicates the CT insulating region. \textbf{d)} The panel displays the $R(\omega)$ of the overdoped sample (blue dots). The black line is the fit of the model dielectric function to the data. The grey area represents the contribution of the CT oscillator to the dielectric function. The left-bottom inset shows the position of the sample in the $p$-$T$ phase diagram.} 
%\label{fig1}
\end{figure}

The ubiquitous instability towards ordered states raises the fundamental question whether these phenomena hide a common and profound origin connected to the existence of an elusive correlated metallic state\cite{Phillips2006,Sordi2011,Phillips2011} that emanates from the Mott insulator and extends up to the critical hole doping level, $p_{cr}\sim$0.16, at which the symmetry-broken orders vanish. 
In charge-transfer systems, such as cuprates, the oxygen bands play a fundamental role in renormalizing the energy scale at which this possible Mott physics can be studied. Considering the simplest case of the parent insulator ($p$=0), the valence fluctuations of Cu-3$d^9$ are suppressed by the strong Coulomb repulsion ($U_{dd}\sim10$ eV) between two electrons occupying the same Cu orbital. The lowest-energy excitation is thus the charge-transfer (CT) of a localized Cu-3$d_{x^2-y^2}$ hole to its neighbouring O-2$p_{x,y}$ orbitals (see Fig. 1a,b), with an energy cost $\Delta_{CT}\sim$2 eV$<U_{dd}$. In the optical conductivity, this process is revealed by a typical CT edge at $\hbar\omega$=$\Delta_{CT}$, which corresponds to the onset of optical absorption by particle-hole excitations in the complete absence of a Drude response\cite{Uchida1991}.

Here we shed new light on the nature of the electronic excitations at the $\Delta_{CT}$ energy scale by adopting a non-equilibrium approach. The high temporal resolution ($\sim$10 fs) of the time-resolved technique employed in this work allows us to access the ultrafast dynamics of the CT excitations before complete thermalization is achieved. We performed experiments on the single-layer Bi$_2$Sr$_{2-x}$La$_x$CuO$_{6+\delta}$ (La-Bi2201) cuprate family (see Methods), in which the hole doping concentration can be accurately controlled by La substitution and can span a broad doping region (0.03$<p<$0.2) across the critical doping $p_{cr}$=0.16. The ultrafast dynamics of the CT excitations is directly compared to the CDW amplitude, that has been measured on the same samples by resonant soft X-ray scattering (RXS) at low temperature\cite{Comin2014}.
These results, supported by dynamical-mean-field-theory (DMFT) calculations, unveil a temperature-independent crossover of the CT dynamics at $p_{cr}\sim$0.16 and suggest that the high-temperature Mott-like state at $p<p_{cr}$ is the necessary precursor of the low-temperature instabilities.

In Figure 1c,d we report the reflectivity curves, $R(\omega)$, for the most underdoped ($p$=0.03, non-superconducting) and the most overdoped ($p$=0.2, $T_c$/$T_{c,max}$=0.57) samples\cite{Lupi2009,Nicoletti2010}. Considering the deeply underdoped sample, the first high-energy optical transition is found at $\Delta_{CT}$=2 eV and hence can be safely ascribed to the charge-transfer process.
When the doping concentration is increased, the energy of this optical transition remains constant (see the inset of Fig. 1c), while its spectral weight progressively decreases. Furthermore, the low-energy region develops a pronounced metallic plasma edge at $\hbar\omega\sim$1 eV.  

Fig. 2a reports the ultrafast dynamics of the CT transition at $T$=300 K in the 1.8-2.5 eV energy range, after the excitation with a 13 fs pulse centered at 1.4 eV. 
%Similar results were obtained with 1.5 eV and 3 eV pump photon energies, thus demonstrating that the pumping process is non-selective and ultimately triggers the creation of a out-of-equilibrium distribution of electron-hole excitations at the CT energy. 
When focusing on the sub-ps dynamics, the data exhibit a clear doping dependence. While the reflectivity variation ($\delta R$($\omega$)/$R$) measured on the underdoped samples is characterized by a pronounced negative (red) signal for $\hbar\omega> 2$ eV, it progressively evolves toward a featureless positive (blue) signal for $p\geq0.16$.

\begin{figure}
\includegraphics[keepaspectratio,clip,width=1\textwidth]{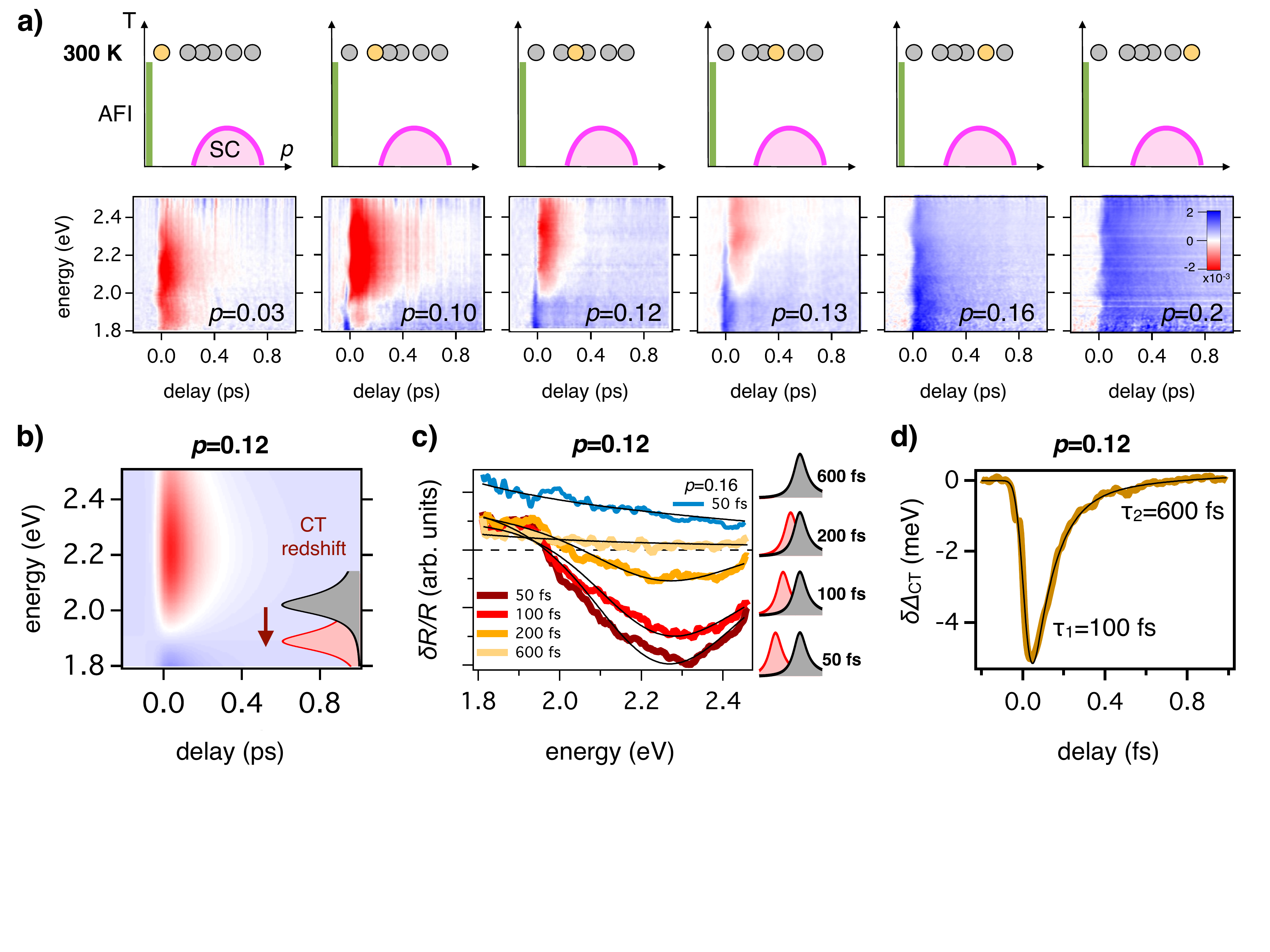}
\caption{\textbf{Ultrafast optical spectroscopy on La-Bi2201}. \textbf{a)} The top row shows the position of the measured samples in the $p$-$T$ phase diagram. In the bottom row we report the $\delta R$($\omega$,$t$)/$R$ maps measured by ultrafast optical spectroscopy on La-Bi2201. The colour scale is reported in the inset. \textbf{b)} Simulation of the $\delta R$($\omega$,$t$)/$R$ signal for the $p$=0.12 sample. A transient redshift of the CT oscillator, described by an exponential decay, is assumed. The colour scale is the same than that in panel a). \textbf{c)} $\delta R$($\omega$)/$R$ spectra at different time delays for the $p$=0.12 sample. The black line is the fit to the data obtained by red-shifting the position of the CT oscillator. For comparison, we report $\delta R$($\omega$)/$R$ for the $p$=0.16 sample (blue line), along with the best fit obtained by increasing the scattering rate in the Drude model. \textbf{d)} Dynamics of $\delta\Delta_{CT}$ for the $p$=0.12 sample. The black line is the fit to the data of a double-exponential decay convoluted with a step function.} 
%\label{fig1}
\end{figure}

The negative $\delta R$($\omega$)/$R$ measured in underdoped samples at $\hbar\omega\sim\Delta_{CT}$ cannot be explained by simply assuming a variation of the total scattering rate of the conduction electorns\cite{DalConte2012,DalConte2015}, since this would lead to a featureless and positive signal over the entire probed frequency range (see Supplementary). In contrast, the $\delta R$($\omega$)/$R$ signal can be perfectly reproduced by assuming a pump-induced redshift of the CT transition (see Supplementary), which results in a reflectivity variation proportional to the derivative of the peak shape. In Fig. 2b we report the  $\delta R$($\omega$,$t$)/$R$ signal calculated by introducing a redshift of the CT peak in the equilibrium dielectric function of the intermediately doped sample ($p$=0.12) and by assuming an exponential decay of the signal. The main features of the experimental transient reflectivity map are qualitatively reproduced by this simple assumption.
For a quantitative analysis of the ultrafast dynamics, we report in Fig. 2c the fit to the $\delta R$($\omega$,$t$)/$R$ spectra for the $p$=0.12 sample at fixed delays ($t$=50, 100, 200, 600 fs), from which we can extract the time evolution of the CT redshift ($\delta\Delta_{CT}$). For all the underdoped samples, the $\delta\Delta_{CT}$ dynamics (see Fig. 2d) is similar and is characterized by two exponential recovery times, $\tau_1\simeq$100 fs and $\tau_2\simeq$600 fs. These timescales are compatible with the coupling to the optical buckling and breathing phonons and, subsequently, to the rest of the lattice vibrations\cite{DalConte2012}, while the coupling of the local charge excitations to short-range antiferromagnetic fluctuations is expected to be effective on the 10 fs timescale\cite{DalConte2015}. The maximum $\delta\Delta_{CT}$ is estimated by considering the value extracted from the fitting procedure at $t$=50 fs. Considering the $p$=0.12 sample, we obtain $\delta\Delta_{CT}$=-5$\pm$1 meV at the excitation density of 7 J/cm$^3$. With this excitation density the maximum value of the CT redshift, i.e., $\delta\Delta_{CT}$=-10$\pm$2 meV, is measured at $p$=0.10 hole doping.

The measured CT redshift discloses important information about the nature of the charge-transfer transition. This process can be easily rationalized starting from the CT insulator ($p$=0), in which the completely localized picture provides a good description of the fundamental electronic excitations. In this framework, the energy necessary to  move a localized hole from the Cu-3\textit{d}$_{x^2-y^2}$ to the O-2\textit{p}$_{x,y}$ orbitals is renormalized by the Coulomb \textit{interatomic} potential ($U_{pd}$) between the excess Cu-3\textit{d} electron and the holes residing on the nearest neighbouring oxygen sites. In simple terms, $U_{pd}$ provides a binding mechanism for the local Cu-3\textit{d}$_{x^2-y^2}-$O-2\textit{p}$_{x,y}$ exciton. Within this local picture, we can sit on a spin-up polarized Cu atom (see Fig. 1a) and assume that the effect of the pump pulse is to transfer to that atom a fraction of spin-down electrons, $\delta \epsilon_{\down}$, from the oxygens within the same CuO$_2$ cell. The excess of positive charges on the oxygen atoms leads to an increase of the binding energy of the additional excitons that can be created on the neighbouring cells by the following probe pulse. This process can be revealed as a decrease of the effective CT energy measured by the probe. Quantitatively, the pump-induced redshift of $\Delta_{CT}$ can be estimated by a simple mean field calculation (see Methods): 
\be
\label{eq:CT_shift}
\delta 	\Delta_{CT} = -\bigg(2U_{pd}
-\fract{5}{24}\,U_{pp}\bigg)\,\big|\delta\epsilon_\down\big|,
\ee
where $U_{pp}$ is the Coulomb repulsion between two charges occupying the same O-2\textit{p} orbital. 
%From a theoretical point of view, the qualitative discussion above strongly suggests that a crucial ingredient to explain the redshift of the charge-transfer gap is the sizable Coulomb interaction, $U_{pd}$, between neighboring copper and oxygen atoms\cite{Hansmann2014}, which is often neglected in theoretical studies of the cuprates. 
Considering the realistic values $U_{pp}\sim$5 eV and $U_{pd}\sim$2 eV (Ref. \citenum{Hansmann2014}) and the photodoping $\delta \epsilon_{\down}\sim$0.3\% (see Methods), we estimate $\delta\Delta_{CT}\sim$-9 meV, which is in very good quantitative agreement with the measured pump-induced redshift in underdoped samples.

Interestingly, the $\delta\Delta_{CT}$ measured in the experiments progressively decreases as the hole doping increases until the $p_{cr}\simeq0.16$ critical doping concentration is reached (see Figure 3a). The $\delta R$($\omega$,$t$)/$R$ signal measured on the optimally ($p$=0.16) and over-doped ($p$=0.2) samples does not show any evidence of a CT redshift, while it can be easily reproduced (see Figure 2c) by assuming an average increase of the electron-boson scattering in the Drude component of the dielectric function, in agreement with the results reported in Refs. \citenum{DalConte2012,DalConte2015}. 
The picture emerging from these results can be summarized as follows: for $p<p_{cr}$, the photoexcitation induces a redshift of the CT transition, which is qualitatively and quantitatively similar to what expected for a CT insulator\cite{Falck1992,Novelli2014}; for $p>p_{cr}$ the ultrafast dynamics can be explained by an increase of the scattering rate of the charge carriers, as expected for a metal. We thus conclude that $p_{cr}$ discriminates, already at high temperature, an underdoped region in which the CT transition is a spatially localized process, as in a Mott insulator, from an overdoped region in which the CT excitation involves wavefunctions spread over many sites, as in more conventional band metals. We note that this localized-delocalized transition of the CT excitation is clearly distinct from the onset of the pseudogap physics, which occurs at a temperature evolving from $T^*\sim$250 K at very low doping to $T^*$=90-150 K at $p$=0.16 and $T^*\sim$50 K at $p$=0.2, as observed by Knight-shift measurements\cite{Kawasaki2010} and confirmed by single-colour pump-probe measurements on the same samples (see Supplementary Information). A similar $T^*(p)$ line has been recently observed by monitoring the $p$-$T$ dependence of the scattering rate of the Drude peak via non-equilibrium infrared spectroscopy in the 0.5-2 eV energy range\cite{Cilento2014}.
Notably, no transition at $p_{cr}$ is observed when we analyse the $\delta R$($\omega$)/$R$ traces extracted at $t>$600 fs (Fig. 2c), i.e., when the excess energy is dissipated in low-energy excitations and converted into heat. This demonstrates that the effect reported here for doped cuprates remains inaccessible to equilibrium techniques, in which only the charge-fluctuations at $k_BT$ are activated. 

\begin{figure}
\includegraphics[keepaspectratio,clip,width=0.6\textwidth]{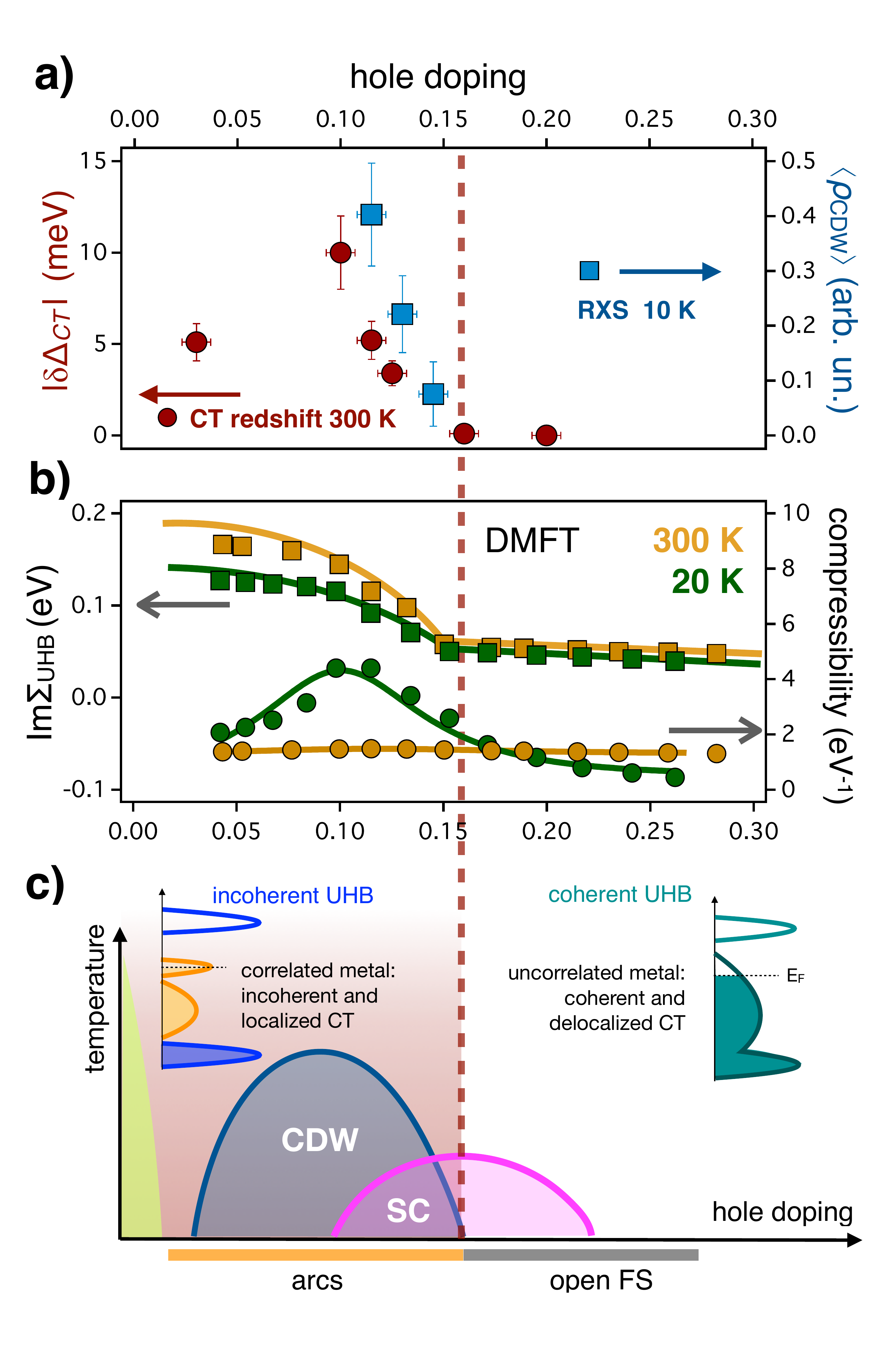}
\caption{\textbf{The high-energy phase diagram of cuprates}. \textbf{a)} The values of the room-temperature CT redshift ($\delta\Delta_{CT}$, red dots, left axis) and the intensity of the low-temperature CDW signal ($\langle\rho_{CDW}\rangle$, blue dots, right axis)  are reported as a function of the hole concentration of the La-Bi2201 samples. The excitation fluence of the pump beam has been tuned in order to maintain a constant absorbed energy density of 7 J/cm$^3$ for the different dopings (see Supplementary). $\langle\rho_{CDW}\rangle$ has been obtained by integrating the difference between the RXS signals measured at the temperatures of 20 K and 300 K and at the CDW wavevector\cite{Comin2014}. The result has been normalized to the total RXS signal at 300 K. Both $\delta\Delta_{CT}$ and $\langle\rho_{CDW}\rangle$ vanish at the critical doping $p_{cr}$=0.16$\pm$0.01, that corresponds to the doping at which a low-temperature transition from Fermi arcs to a closed Fermi surface has been measured by STM\cite{He2014}. \textbf{b)} The imaginary part of the electronic self-energy (colored squares) is calculated by DMFT and is reported as a function of the doping concentration for different temperatures (300K dark yellow; 20 K green). The calculated electronic compressibility is indicated by colored circles. The full lines are guides to the eye. \textbf{c)} A sketch of the non-equilibrium $p$-$T$ phase diagram of copper oxides is reported. The pink, blue and green areas delimit the superconducting (SC) dome, the charge-ordered (CDW) state and the antiferromagnetic insulator, respectively.} 
%\label{fig1}
\end{figure}

The crucial idea that drives the present work is that the Mott-like nature of the electronic states, that we probe at the energy scale $\Delta_{CT}$, is the fundamental prerequisite for the development of low-temperature instabilities.
A possible link between the high- and low-energy physics is that a correlated metal in proximity of the Mott insulating phase is characterized by a reduced mobility of the charge carriers confined into a narrow band at the Fermi level. Upon small variations of the chemical potential $\mu$, the density of states at the Fermi level is expected to dramatically change, thus possibly leading to a very high electronic compressibility\cite{Grilli1991,Sordi2012}, $K\propto \partial n/\partial \mu$. At low temperatures the additional freezing of the thermal excitations renders the system naturally prone to phase separation, in which strong charge inhomogeneities ($\delta n$) can coexist at the same chemical potential.
DMFT calculations provide a solid support to this naive picture in the case of a realistic three-band model with interaction terms among the O-2$p_{x,y}$ and Cu-d$_{x^2-y^2}$ orbitals similar to those used in Eq. \ref{eq:CT_shift} (see Methods). In particular, we focus on the doping-dependence of the upper Hubbard band (UHB), that corresponds to the double occupation of the Cu-3\textit{d}$_{x^2-y^2}$ levels, i.e., the final state of the CT process. In Figure 3b we plot the imaginary part of the electronic self energy, that corresponds to the inverse lifetime, of the UHB (Im$\Sigma_{UHB}$). At temperatures as high as 300 K the large value of the inverse lifetime, typical of local incoherent excitations in the vicinity of the Mott insulating state, progressively decreases until the $p_{cr}$ doping is reached. Above this value, Im$\Sigma_{UHB}$ shows a smaller and almost constant value that indicates the transition to delocalized and coherent excitations, similar to what expected for conventional interband transitions in uncorrelated solids. In the three-band model considered, the onset of coherence of the CT excitations at $p_{cr}$ coincides with the merging of the quasiparticle peak at the Fermi level with the broad conduction band of mixed $p$-$d$ character, characteristic of the doping-driven Mott transition in DMFT \cite{Fisher1995}. As expected, the computed high-temperature compressibility is a smooth function of doping and does not evidence any tendency to charge-separation (Figure 3b). The picture dramatically changes at low temperature, when the additional freezing of the charge carriers in the correlated states close to $\mu$ cooperates in increasing $K$ in the underdoped region of the phase diagram. As shown in Figure 3b, while the high-energy Im$\Sigma_{UHB}$ transition is almost unaffected by temperature, the compressibility shows at low temperatures a pronounced maximum for $p<p_{cr}$ that suggests the tendency to develop charge inhomogeneities. Even though the emergence of CDW at a specific wavevector is the result of more complex ingredients, such as the long-range Coulomb interactions\cite{Castellani1995} and the topology of the Fermi surface, our results suggest that the proximity to the Mott-state is the prerequisite for the low-temperature development of charge-order instabilities.         

The validity of this picture is further corroborated by the outcome of RXS measurements at $T$=10 K on the same samples, as previously measured in Ref. \citenum{Comin2014}. 
The spontaneous breaking of the translational symmetry of the charge distribution within the CuO$_2$ planes is detected as a resonance in the RXS signal at a specific exchanged parallel wavevector (see Supplementary Information). While the width of the RXS peaks indicates a CDW correlation length of the order of 2-3 nm, the momentum-integrated signal can be taken as proportional to the average amplitude of the charge-density modulation. Fig. 3a reports the CDW amplitude, $\langle\rho_{CDW}\rangle$, on La-Bi2201 at different hole doping concentrations, obtained by integrating the RXS signal measured at the proper CDW wavevector. The CDW amplitude progressively decreases until $\langle\rho_{CDW}\rangle\rightarrow0$ at $p_{cr}$=0.16$\pm$0.01, that is the same doping concentration at which the $\delta\Delta_{CT}$ signal vanishes. This observation undoubtedly demonstrates that the development of short-ranged CDW at low temperature takes place only in the doping region $p<p_{cr}$, that is characterized, already at room temperature, by the Mottness of the UHB.

Taken together, these observations consistently show that the phase diagram of copper oxides is characterized by a temperature-independent transition from a correlated to a more conventional metal at $p_{cr}\simeq$0.16. This phenomenon, that involves energy scales as high as $\Delta_{CT}$, can be observed already at room temperature via non-equilibrium experiments.
Our observations impact on many aspects of the physics of copper oxides: i) charge-order emerges as the low-energy manifestation of a correlated ground state in the $p<p_{cr}$ region of the phase diagram; ii) while low-energy models, which take into account the details of the electronic interactions at the Fermi level, are necessary to correctly predict the CDW wavevector, symmetry and onset temperature, the value $p_{cr}$ at which the charge order vanishes is the consequence of a high-energy phenomenon; iii) any theory for the charge-order phenomenon should rely on the correlated nature of the electronic states at the Fermi level, which is reflected in the quenching of the O-2\textit{p}$\rightarrow$Cu-3\textit{d} charge fluctuations at the energy scale $\Delta_{CT}$ and in the freezing of the charge carriers that drives the upturn of the electronic compressibility at $p<p_{cr}$.

More in general, we note that the critical doping $p_{cr}$ is a turning point for many low-temperature properties of copper oxides, such as the momentum space topology\cite{Fujita2014,He2014}, the ARPES quasiparticle strength\cite{Fournier2010}, the superconductivity-induced kinetic energy change\cite{Deutscher2005,Giannetti2011}, the time-reversal symmetry breaking\cite{Xia2008}, the change of the in-plane resistivity curvature\cite{Ando2004}, the transition from $p$ to $1+p$ charge carrier density\cite{Ono2000,Badoux2016,Laliberte2016}, the crossover of spin excitations from damped spin-waves to incoherent spin-flips\cite{Minola2016} and the strong increase of the quasiparticle effective mass\cite{Ramshaw2015}. 
Our results suggest a novel intriguing scenario, in which the crossover at $p_{cr}$  between the physics of a doped Mott insulator to that of a more coherent metal is at the origin of the low-temperature phenomenology. In this framework, the reduced mobility of the charge carriers associated with the Mottness for $p<p_{cr}$ constitutes the fertile ground for the onset of lower-symmetry instabilities which are generally attributed to a putative $T$=0 quantum critical point hidden by the superconducting state.

\section{Methods}
\subsection{Experiments}
A Ti:sapphire amplifier (Clark-MXR model CPA-1) delivers a train of pulses at 1 kHz repetition rate with 150-fs duration at 780 nm central wavelength and is used to simultaneously drive two Non-collinear Optical Parametric Amplifiers (NOPAs) operating in different frequency intervals. All NOPAs are seeded by white light continuum (WLC) generated in a sapphire plate. The first NOPA (NOPA1) is pumped by the second harmonic and amplifies in a beta-barium borate (BBO) crystal pulses with a spectral content between 820 nm (1.5 eV) and 1050 nm (1.2 eV), which are compressed to nearly TL 13-fs duration by a couple of fused silica prisms. This NOPA serves to trigger the dynamics and it is synchronized with a second NOPA  (NOPA2), pumped by the second harmonic and using BBO, which is used to probe the reflectivity variation of the system. The spectrum of NOPA2 spans a frequency range between 510 nm (2.4 eV) and 700 nm (1.8 eV) and it is compressed to 7 fs duration by multiple bounces on a pair of chirped mirrors, making the overall temporal resolution of the pump-probe setup below 15 fs. The time delay between pump and probe is adjusted by a motorized delay stage and both the beams are focused on the sample by a spherical mirror in a quasi-collinear geometry. The reflected probe spectrum is detected by a Si spectrometer working at the full 1 kHz laser repetition rate. By recording the reflected probe spectrum at different temporal delays t with and without pump excitation, we measure the differential reflectivity: $\delta R(\omega,t)/R(\omega)$=$[R(\omega,t)$-$R_{eq}(\omega)]/R_{eq}(\omega)$. The pump fluence used for the experiments is 500 $\mu$J/cm$^2$. The density of CT excitations can be estimated starting from the pump penetration length ($l_{pen}$) at a specific doping (see Supplementary Information). For example, assuming $l_{pen}\sim$700 nm for $\hbar\omega$=1.4 eV and $p$=0.10 we obtain an absorbed energy of $\sim$7 J/cm$^3$, which corresponds to a density of $\Delta_{CT}$=2 eV excitations of about 2x10$^{19}$ cm$^{-3}$. Considering that the density of Cu atoms is $\sim$6x10$^{21}$ cm$^{-3}$, we obtain that the fraction of holes transferred from the Cu atoms is $\delta \epsilon_{\down}$ $\sim$3x10$^{-3}$.
The La-Bi2201 crystals were grown using the floating-zone technique, and characterized as described in Ref. \citenum{Ono2003}. The doping has been determined following Ref. \citenum{Ando2000}.

%The scattering measurements were performed at beamline UE46 in BESSY and beamline REIXS at the Canadian Light Source, using a 3- and a 4-circle diffractometer, respectively, using vertically polarized incoming light. In order to maximize the charge order signal, all measurements were taken at the peak energy of the Cu-$L_3$ edge (931.5 eV), and at $T$=10 K.

\subsection{Mean-field calculation of the CT redshift}
In the fully atomic picture (half-filling), the hamiltonian governing the physics of the CuO$_2$ plane can be written as:
\ba
H&=&\epsilon_\mathrm{Cu}\,n_d + \epsilon_\mathrm{O}\,\Big(n_{\mathrm{O}_1} + n_{\mathrm{O}_2}\Big) + \fract{U_{dd}}{2}\,\big(n_d-1\big)^2 + \fract{U_{pp}}{2}\,\sum_{i=1,2} \Big(n_{\mathrm{O}_i}-6\Big)^2\\
&& + U_{pd}\,\big(n_d-1\big)\,\sum_{i=1,2} \Big(n_{\mathrm{O}_i}-6\Big)
\ea
in order to minimize the interaction when one electron sits on the Cu atom and both the oxygens (labeled by $i$) are fully occupied, i.e., $\langle n_d + n_\mathrm{O_1} + n_\mathrm{O_2}\rangle$=13. As a crude approximation, we assume a localized spin (up) on the Cu atom through the following parametrization: $\langle n_{d\up} \rangle$=$1-\epsilon_\up,$, $\langle n_{d\down} \rangle$=$-\epsilon_\down,$ and $\langle n_{{\mathrm{O}_1}\up(\down)}\rangle$=$\langle n_{\mathrm{O_2}\up(\down)}\rangle$=$3+\fract{\epsilon}{2}$, where $2\epsilon=\epsilon_\up+\epsilon_\down$ is the total photoinduced change of the occupation of the Cu$_{\up}$ and Cu$_{\down}$ sites.

The mean-field calculation of the mean value of the Cu and O levels, i.e., $\mu_{\mathrm{Cu},\sigma}$ and $\mu_{\mathrm{O},\sigma}$, result to be:
\ba
\mu_{\mathrm{Cu},\up(\down)} &=&  \epsilon_\mathrm{Cu} - U_{dd}\,\big(\epsilon_{\down(\up)}+\small(-\small)1/2\big) + 2U_{pd}\epsilon\\
\mu_\mathrm{O} &=& \epsilon_\mathrm{O} + \fract{5}{24}\,U_{pp}\,\Big(2\epsilon-\frac{72}{5}\Big)
- 2\,U_{pd}\,\epsilon. 
\ea
In the simplest case of a single band model, the difference between the empty and occupied Cu levels, i.e., $\mu_{\mathrm{Cu},\down}-\mu_{\mathrm{Cu},\up}$=$U_{dd}$, results independent of the photoinduced occupation, since intrinsically $\epsilon_\up$=$\epsilon_\down$. This result suggests that in the single-band Mott insulator, the density of states of the upper- (UHB) and lower-Hubbard bands (LHB) decreases upon photoexcitation, while the gap value remains constant. 

The scenario is qualitatively different for a charge-transfer insulator, in which the electrons are transferred from the Cu to the O atoms and the $\epsilon_\up$=$\epsilon_\down$ symmetry is broken. If we sit on the spin up Cu site, the effect of the pump excitation is to transfer a certain amount of electrons from the oxygen to the Cu spin down state, i.e., $\delta\epsilon_\down<0$ while $\delta\epsilon_\up=0$.
Therefore, the change of the charge-transfer gap measured by the probe pulse is given by:
\be
\delta \big(\mu_{\mathrm{Cu},\down}-\mu_\mathrm{O}\big) = -\bigg(2U_{pd}
-\fract{5}{24}\,U_{pp}\bigg)\,\big|\delta\epsilon_\down\big|,
\ee

\subsection{Mean field theory calculations}
We consider a model including copper $d_{x^2-y^2}$ orbitals and oxygen $p_x$ and $p_y$ orbitals with the same interaction terms as in the Hartree-Fock calculation and near neighbor hopping between copper and oxygen and oxygen-oxygen. The parameters are $U_{dd}$=10 eV, $U_{pp}$=5 eV, $U_{pd}$=2 eV, $\Delta_{CT}$=2 eV, $t_{pd}$=0.3 eV, $t_{pp}$=0.1 eV. We solve the model using single-site DMFT treating both the copper-oxygen and oxygen-oxygen repulsions at the Hartree-Fock level, while the copper-copper interaction is included without approximations.
The impurity model is solved using exact diagonalization at finite temperature with 8 levels in the bath ($Ns$=9) and keeping 50 states in the calculation of the trace. We have verified that the results for statical observables are converged in both truncation parameters.

\bibliography{QCP}

%% Here is the endmatter stuff: Supplementary Info, etc.
%% Use \item's to separate, default label is "Acknowledgements"

%\item We thank M. Grilli, F. Cilento, G. Coslovich, D. Fausti, F. Parmigiani for the useful and fruitful discussions. We gratefully acknowledge D. Bonn and B. Keimer for the support in the development of the MPI-UBC Tl2201$_{\mathrm{OD}}$ research effort.
\section*{Acknowledgments} The research activities of M.F. have received funding from the European Union, under the project ERC-692670 (FIRSTORM). F.B acknowledge financial support from the MIUR-Futuro in ricerca 2013 Grant in the frame of the ULTRANANO Project (project number: RBFR13NEA4). 
MC acknowledges funding by SISSA/CNR project "Superconductivity, Ferroelectricity and Magnetism in bad metals" (Prot. 232/2015).
F.B., G.F. and C.G. acknowledge support from Universit\`a Cattolica del Sacro Cuore through D1, D.2.2 and D.3.1 grants. F.B and G.F acknowledge financial support from Fondazione E.U.L.O. D.B. acknowledges the Emmy Noether Programm of the Deutsche Forschung Gemeinschaft. G.C. acknowledges funding from the European Union Horizon 588 2020 Programme under Grant Agreement 696656 Graphene 589 Core 1.
%L.V. is supported by the Alexander von Humboldt Foundation.
%M.M. acknowledges support from the NCN project DEC-2013/09/B/ST3/01659.
%The Y-Bi2212$_{\mathrm{UD}}$ crystal growth work was performed in M.G.Õs prior laboratory at Stanford University, Stanford, CA 94305, USA, and supported by the US Department of Energy, Office of Basic Energy Sciences.
%The work at UBC was supported by the Max Planck - UBC Centre for Quantum Materials, the Killam, Alfred P. Sloan, Alexander von Humboldt, and NSERC's Steacie Memorial Fellowships (A.D.), the Canada Research Chairs Program (A.D.), NSERC, CFI, and CIFAR Quantum Materials.
%M.C. is financed by European Research Council through FP7/ERC Starting Grant SUPERBAD, Grant Agreement 240524.
%J.B. acknowledges support by the P1-0044 of ARRS, Slovenia.
%G.C. acknowledges support by the EC under Graphene Flagship (contract no. CNECT-ICT-604391).
%N.D.Z. acknowledges support from the NCCR project Materials with Novel Electronic Properties and appreciates expert collaboration with J. Karpinski.

%%
%% TABLES
%%
%% If there are any tables, put them here.
%%
\section{Supplementary Material}
\begin{figure}[h]
\centering
\includegraphics[width=10cm]{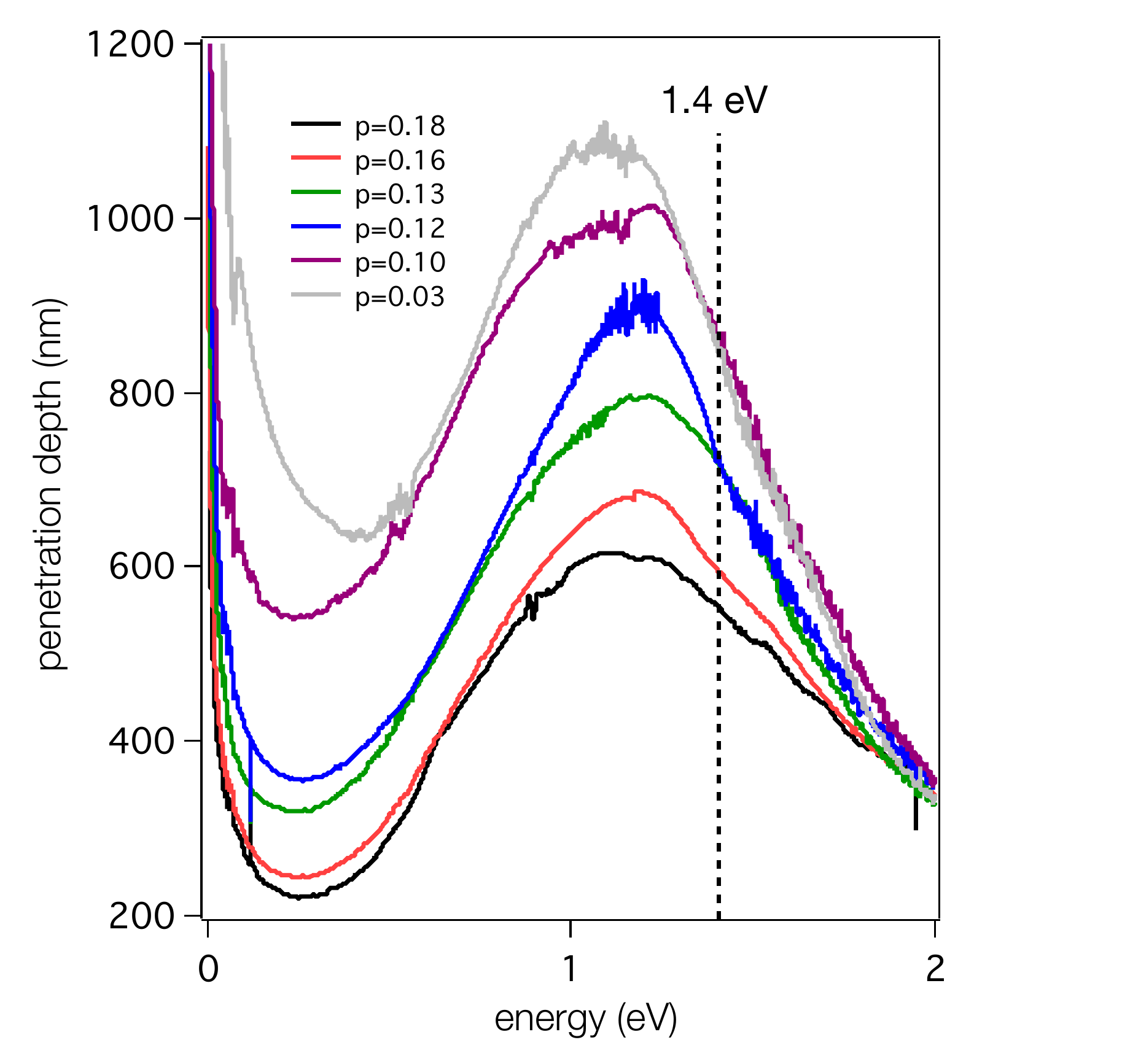}
\caption{Penetration depth in $\mathrm{Bi}_2\mathrm{Sr}_{2-x}\mathrm{La}_x\mathrm{CuO}_{6+\delta}$, as a function of the photon energy and for different doping concentrations, $0.03<p<0.18$. The photon energy of the pump pulse (1.4 eV) is indicated by a black line.}
\label{fig:Sigma_PenDepth}
\end{figure}
\subsection{Optical Properties of $\mathrm{Bi}_2\mathrm{Sr}_{2-x}\mathrm{La}_x\mathrm{CuO}_{6+\delta}$ at equilibrium}
In Figure S\ref{fig:Sigma_PenDepth} we report the light penetration depth in La-Bi2201, as a function of the photon energy at different doping concentrations. The penetration depth ($d_{pen}$) has been obtained from the La-Bi2212 optical conductivity that has been measured elsewhere \citep{Lupi2009}. Since the density of the CT excitations induced by the pump pulse depends on the energy density delivered by the pump pulse, the knowledge of $d_{pen}$ is necessary to maintain a constant excitation density when the doping is changed. As shown in Figure S1, at the pump photon energy of 1.4 eV $d_{pen}$ decreases as the doping is increased. This is the consequence of the progressive increase of the charge carrier density which results in a larger Drude contribution to the absorption process. The incident fluence has been tuned in order to maintain a constant excitation density of 7 J/cm$^3$ for all the time-resolved measurements at different doping concentrations. \ignorespaces

\subsection{Ultrafast reflectivity variation in the UV-vis spectral range}
\begin{figure}[h]
\centering
\includegraphics[width=7cm]{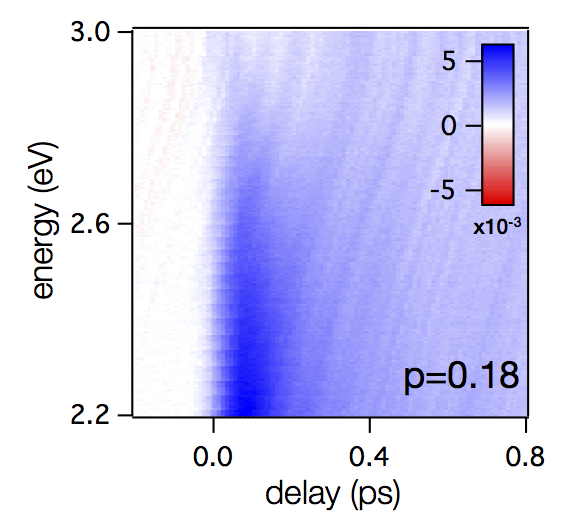}
\caption{Ultrafast reflectivity variation ($\delta R(\omega,t)/R$) measured by optical spectroscopy on the $p$=0.2 La-Bi2201 sample. The pump photon energy was set at 2.06 eV. The colour scale is reported in the inset.}
\label{fig:x02_UV}
\end{figure}

In order to rule out the possibility that, in optimally and over-doped La-Bi2201 samples, the absence of the CT redhsift is related to a large increase of the $\Delta_{CT}$ energy, which would push the observed phenomenon out of the explored energy window, we extended the time-resolved measurements up to an energy of 3 eV. In Figure S\ref{fig:x02_UV} we report the $\delta R(\omega,t)/R$ matrix measured on the overdoped La-Bi2201 sample ($p$=0.2). The absence of any negative component in the signal, allows us to conclude that no CT redshift is observed in the energy range that extends far above the $\Delta_{CT}$ energy estimated by equilibrium optical spectroscopy. The same experiment has been repeated at different pump photon energies ($\hbar\omega=2.06$ eV and $\hbar\omega=1.77$ eV). The measured dynamics did not evidence any significant change, in agreement with the results reported in the main text.

\begin{figure}[h]
\centering
\includegraphics[width=14cm]{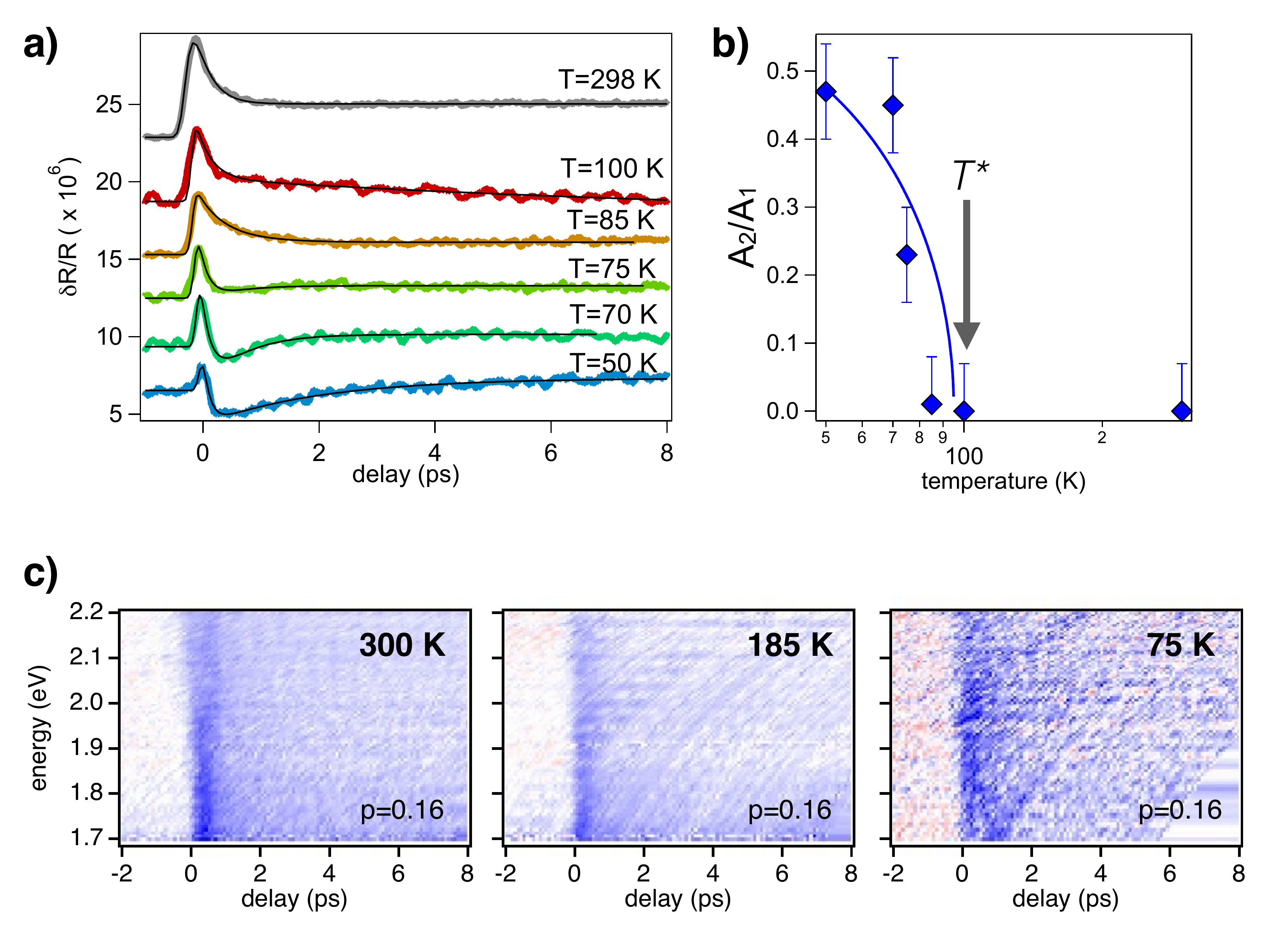}
\caption{a) Single-color ultrafast reflectivity variation ($\delta R(t)/R$) maesured at different temperatures on the $p$=0.16 La-Bi2201 sample. The solid black lines represent the bi-exponential fit to the data. b) The ratio ($A_2/A_1$) of the amplitude of the two exponential functions is reported as a function of the temperature. The solid line is a guide to the eye. c) Ultrafast dynamics at the $\Delta_{CT}$ energy scale at different temperatures. The color scale is the same than that used in Figure S\ref{fig:x02_UV}}
\label{fig:pseudogap}
\end{figure}\ignorespaces
\subsection{Ultrafast optical spectroscopy at different temperatures}
The transition of the CT dynamics observed in the La-Bi2201 samples at room temperature and at $p$=$p_{cr}$ is uncorrelated with the onset of the pseudogap at low temperatures. This can be easily inferred by single-color reflectivity measurements on the $p$=0.16 La-Bi2201 sample at different temperatures, as shown in Figure S\ref{fig:pseudogap}. The data have been collected starting from a cavity-dumped Ti:Sapphire oscillator. The photon energies are set to 3.14 and 1.5 eV for the pump and the probe, respectively. The fluence is of the order of 10 $\mu$J/cm$^2$. Panel a) shows the time traces at different temperatures. The onset of a negative component, typical of the pseudogap phase \cite{Cilento2014} is observed below a temperature $T^*\simeq$100 K. A double exponential function $\delta R(\tau)/R=A_1e^{\frac{-t}{\tau_1}}+A_2e^{\frac{-t}{\tau_2}}$ is fit to the measured time traces. While the first exponential is positive and accounts for the typical electron-phonon dynamics of the normal state \cite{DalConte2012}, the amplitude of the second decay $A_2$ is negative and accounts for the dynamics in the pseudogap region \cite{Cilento2014}. In Figure S\ref{fig:pseudogap} we report the absolute value of the ratio between the negative and the positive contributions ($A_2/A_1$) as a function of the temperature. The $A_2$ component vanishes at $T^*\simeq$100 K demonstrating that the pseudogap onset is at temperatures significantly smaller that the temperature at which the $p=p_{cr}$ discontinuity is observed.

The Mottness in the $p<p_{cr}$ region of the phase diagram involves energy scales corresponding to $\Delta_{CT}$=2 eV. The dynamics at such a high energy scale is expected to be completely temperature independent, being the thermal fluctuations confined to $k_BT$. In order to support this assumption, we performed frequency- and time-resolved measurements at different temperatures. In particular, we focused on the optimally doped sample that is the closest to the $p_{cr}$ turning point. In order to avoid artifacts related to the impulsive heating of the sample when the temperature is decreased we performed time-resolved measurements in the low-fluence regime ($\simeq$10 $\mu$J/cm$^2$). The relative reflectivity variation, $\delta R(\omega,t)/R$, has been measured exploiting the supercontinuum light produced by a photonic fiber seeded by a cavity-dumped Ti:sapphire oscillator. The details of the experimental setup can be found in Refs. \citenum{Giannetti2011} and \citenum{Cilento2009} The frequency- and time-resolved reflectivity maps are reported in Figure S\ref{fig:pseudogap}c. In order to avoid effects related to the increased average heating at low temperature, the repetition rate (RR) of the experiment has been decreased as to maintain the ratio RR/$C_{tot}(T)$ constant, $C_{tot}(T)$ being the total heat capacity. The data reported in the figure demonstrate that the $\delta R(\omega,t)/R$ signal at the $\Delta_{CT}$ energy scale is temperature independent and that the transition observed at $p_{cr}$ does not represent the room-temperature intersection with an additional $\tilde{T}(p)$ line that decreases as the doping increases.

\section{Differential analysis of the transient reflectivity variation induced by changes in the equilibrium dielectric function}
\begin{figure}[h]
\centering
\includegraphics[width=13cm]{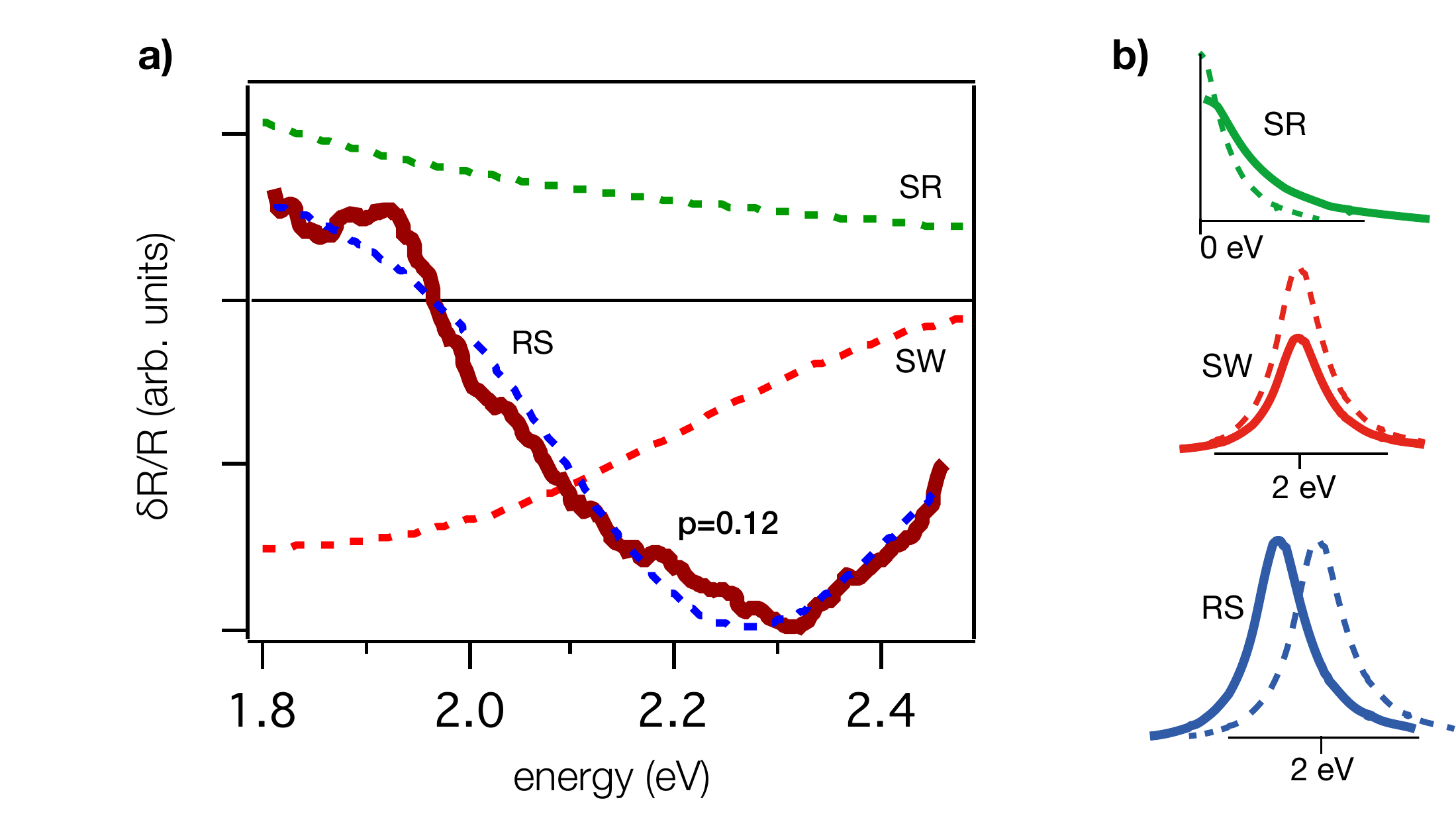}
\caption{a) $\delta R(\omega,\tau)/R$ (purple solid line) measured on the $p$=0.12 La-Bi2201 sample at a delay $\tau$=50 fs. The dashed lines are the reflectivity variations calculated by modifying different parameters in the   equilibrium dielectric function. In particular, we considered the increase of the total scattering rate in the Drude part of the optical conductivity (dashed green line, SR), the redshift of $\Delta_{CT}$ (dashed blue line, RS) and the change of the CT spectral weight (dashed red line, SW). b) A pictorial view of the modification of the optical conductivity for the three cases is reported.}
\label{fig:ParReflVar}
\end{figure}\ignorespaces

The analysis of the time-resolved data has been carried out starting from the equilibrium dielectric function of the samples that has been measured elsewhere \cite{Lupi2009}. The best fitting to the complex optical conductivity has been obtained by combining a Drude model\cite{DalConte2012,DalConte2015} and high-energy Lorentz oscillators:
\begin{eqnarray}\label{eq:Drude_Lorentz_Model}
\sigma(\omega)=\dfrac{1}{4\pi}\dfrac{\omega^2_{pD}}{1/\tau_D-i\omega}+\dfrac{\omega}{4\pi}\sum_j\dfrac{\omega^2_{pj}}{\omega /\tau_j-i(\omega^2-\omega_j^2)}
\end{eqnarray}
The first term refers to the relaxation of the free charge carriers with the scattering rate $\gamma=1/\tau_D$; the second term is a sum of Lorentz oscillators - characterized by the central frequency $\omega_j$, the strength of the oscillator $\omega^2_{pj}$ and the scattering rate $\gamma_j=1/\tau_j$ - that describe the response of bound charges. 

The idea of the differential model is to find the minimum number of parameters in the equilibrium dielectric function which have to be modified to reproduce the reflectivity variation, i.e., $\delta R(\omega,\tau)/R$, measured at a given time delay $\tau$.
As discussed in Refs. \citenum{DalConte2012} and \citenum{DalConte2015}, the $\delta R(\omega,\tau)/R$ signal measured on optimally and overdoped copper oxides can be interpreted, already after $\sim$40 fs, as a transient increase of the electron-boson scattering rate. The increase of the scattering rate induces a broadening of the Drude plasma edge across the plasma frequency at $\omega_D=\sim$1 eV. The $\delta R(\omega,\tau)/R$ detected at probe frequencies $\omega>\omega_D$ results in a positive and featureless signal, which monotonically decreases at high frequencies. This behaviour, that is confirmed in the measurements on La-Bi2201 for hole concentrations $p\geq$0.16 , clearly contrasts with the $\delta R(\omega,\tau)/R$ signal observed at $p<0.16$ and for $\tau \lesssim$600 fs (see Fig. 2 of the main manuscript). As an example, we report in Fig. S\ref{fig:ParReflVar} the signal $\delta R(\omega,50 fs)/R$ measured on the underdoped La-Bi2201 sample with $p$=0.12. Clearly, the negative reflectivity variation at $\omega>\Delta_{CT}\simeq$2 eV cannot be attributed to a change of the electron-boson scattering rate (dashed green line, SR). On the other hand, the $\delta R(\omega,\tau)/R$ signal is perfectly reproduced simply by assuming a redshift of the CT oscillator (dashed blue line, RS) alone. For completeness, we also show that a change in the spectral weight of the CT oscillator (dashed red line, SW) does not account for the measured $\delta R(\omega,\tau)/R$.   

\subsection{Resonant soft X-ray scattering (RXS) on $\mathrm{Bi}_2\mathrm{Sr}_{2-x}\mathrm{La}_x\mathrm{CuO}_{6+\delta}$}
\begin{figure}[h]
\centering
\includegraphics[width=13cm]{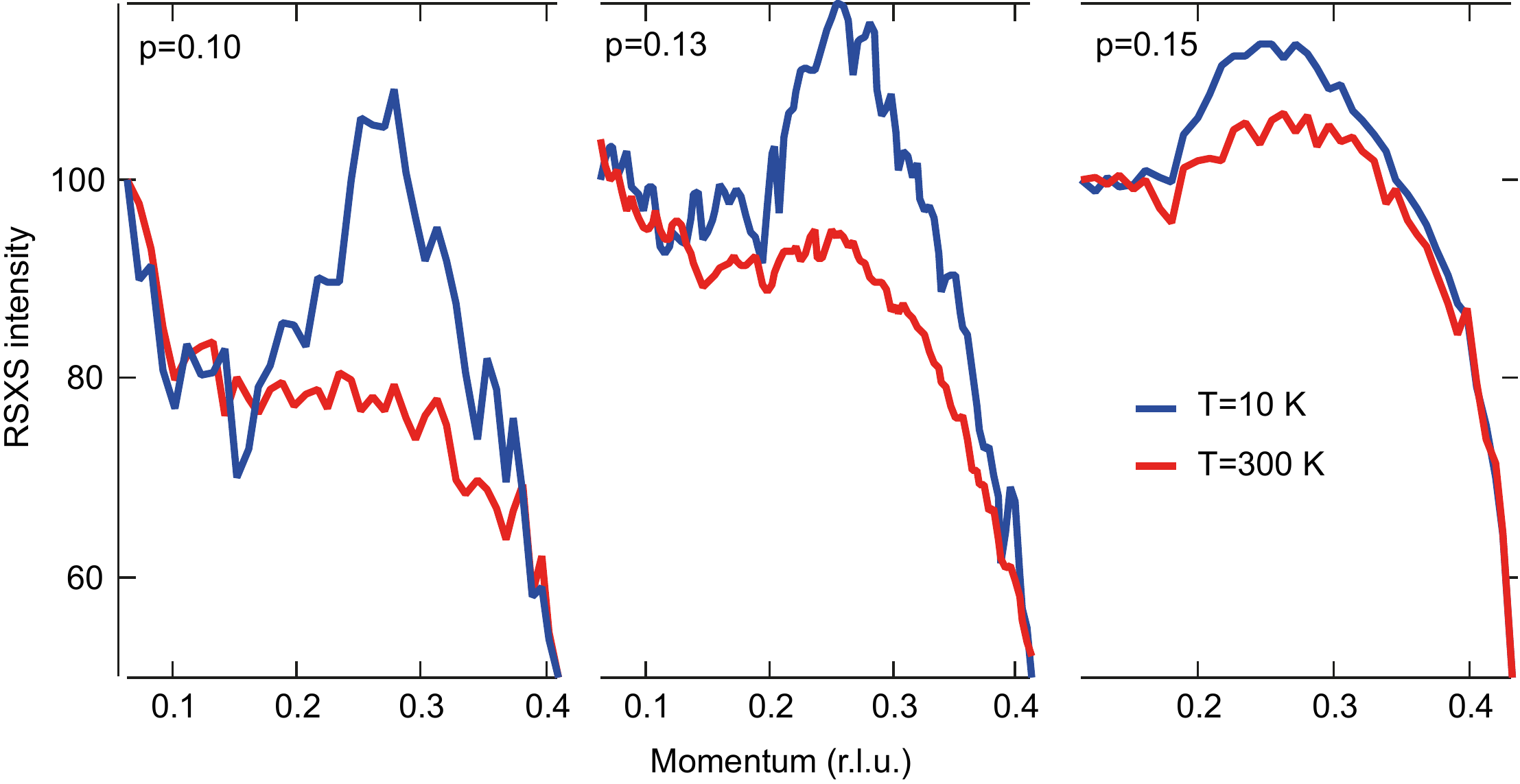}
\caption{Low- and high-temperature scattering scans for the doping levels investigated with RXS: $p=0.10$, $p=0.13$, $p=0.15$ \cite{Comin2014}.}
\label{fig:RXS}
\end{figure}\ignorespaces
In Fig. S\ref{fig:RXS} we show resonant soft X-ray scattering (RXS) measurements on $\mathrm{Bi}_2\mathrm{Sr}_{2-x}\mathrm{La}_x\mathrm{CuO}_{6+\delta}$ at three different level of doping. These data are readapted from \cite{Comin2014}. The resonance in the RXS signal at T=10 K at momentum $\mathbf{Q}_{\|}=0.27$ underlines the presence of a spontaneous breaking of the translational symmetry of the charge distribution within the CuO$_2$ planes. The amplitude of the charge density modulation ($\langle\delta_{CDW}\rangle$) shown in Fig. 3 of the main text for different doping is evaluated integrating the RXS signal along the whole probed momenta at T=10 K and normalizing it on the integrated intensity at T=300 K.

%\bibliography{PhDThesisBib} % Use the bibliography.bib file for the bibliography
%\bibliographystyle{plainnat}

\end{document}